\documentclass[12pt]{article}
\begin{document}

\begin{center}
{\Large{BRST charge for nonlinear algebras}}\\
\vspace{1cm}

{\large I.L. Buchbinder}\footnote{E-mail:joseph@tspu.edu.ru},
{\large P.M.~Lavrov}\footnote{E-mail: lavrov@tspu.edu.ru}

\bigskip
{\normalsize\it {Tomsk State Pedagogical University,
634041 Tomsk, Russia}}

\end{center}

\thispagestyle{empty}

\begin{abstract}\noindent
We study the construction of the classical nilpotent canonical BRST
charge for the nonlinear gauge algebras where a commutator (in terms
of Poisson brackets) of the constraints is a finite order polynomial
of the constraints.

\end{abstract}

\section{Introduction}

The BRST charge, corresponding to the Noether current of BRST symmetry \cite{brst}, is
one of the most efficient tools for studing the classical and
quantum aspects of constrained systems. The
properties of the BRST charge, especially its nilpotency, are the
base of modern quantization methods of gauge theores in both
Lagrangian \cite{BV} and Hamiltonian \cite{BFV} formalism.

In this paper we discuss the form of the canonical BRST charge for a general
enough class of gauge theories. Classical formulation of the gauge theory
in phase space is characterized by first class constraints
$T_{\alpha}=T_{\alpha}(p,q)$ with $p_i$ and $q^i$ being canonically
conjugate phase variables. Constraints $T_{\alpha}$ satisfy the
involution relations in terms of the Poisson bracket
\begin{eqnarray}\label{i0}
\{T_{\alpha},T_{\beta}\}=f^{\gamma}_{\alpha\beta}T_{\gamma}
\end{eqnarray}
with structure functions $f^{\gamma}_{\alpha\beta}$. In
Yang-Mills type theories the structure functions are constants and
the nilpotent BRST charge ${\cal Q}$ ($\{{\cal Q},{\cal Q}\}=0$) can
be written in a closed form.
For general gauge theories the structure functions
depend on phase variables
$f^{\gamma}_{\alpha\beta}=f^{\gamma}_{\alpha\beta}(p,q)$ and the
existence theorem for the nilpotent BRST charge has been proved
\cite{BF}. It allows to present ${\cal Q}$ by series expansion (in
general, infinite) in ghost variables
\begin{eqnarray}\label{i2}
{\cal Q}=c^{\alpha}T_{\alpha}-\frac{1}{2}c^{\alpha}c^{\beta}
f^{\gamma}_{\alpha\beta}{\cal P}_{\gamma}+\cdot\cdot\cdot =
{\cal Q}_1+{\cal Q}_2+\cdot\cdot\cdot.
\end{eqnarray}
Here $c^{\alpha}$ and ${\cal P}_{\alpha}$ are canonically conjugate
ghost variables and the dots mean the terms of higher orders in ghost variables
conditioned by $p,q$ dependence of structure functions.  The
problem, which we discuss here, consists in construction for a given
constrained theory the higher order contributions to ${\cal Q}$ in
terms of its structure functions. In general, solution to this
problem is unknown.

We consider a class of gauge theories which can be described
in terms of  constraints
$T_{\alpha}$ satisfying the relation (\ref{i0}) with nonconstant
structure functions which form a finite order polynomial in the
constraints $T_{\alpha}$
\begin{eqnarray}\label{j}
f^{\gamma}_{\alpha\beta}= F^{\;\;\gamma}_{\alpha\beta}+
V^{(1)\gamma\beta_1}_{\alpha\beta}T_{\beta_1}+\cdot\cdot\cdot+
V^{(n-1)\gamma\beta_1...\beta_{n-1}}_{\alpha\beta}T_{\beta_1}\cdot\cdot\cdot
T_{\beta_{n-1}}
\end{eqnarray}
where $F^{\;\;\gamma}_{\alpha\beta},
V^{(1)\gamma\beta_1}_{\alpha\beta},...,
V^{(n-1)\gamma\beta_1...\beta_{n-1}}_{\alpha\beta}$ are constants.
Algebras of such kind appeared to be in conformal field theories
(the so-called ${\cal W}_N$ algebras) \cite{CFT}, in theories with
quantum groups \cite{IO}, in higher spin theories on AdS space \cite{HSF}.

Construction of the nilpotent BRST charge for quadratically
nonlinear algebras ($ V^{(2)}=...=
V^{(n-1)}=0$) subjected to
an additional special assumption concerning structure constants
$V^{(1)\gamma\delta}_{\alpha\beta}=V^{\gamma\delta}_{\alpha\beta}$
 was performed in \cite{SSvN} with the result
\begin{eqnarray}\label{i3}
{\cal Q}=c^{\alpha}T_{\alpha}-\frac{1}{2}c^{\alpha}c^{\beta}
F^{\gamma}_{\alpha\beta}{\cal P}_{\gamma}-
\frac{1}{2}c^{\alpha}c^{\beta}V^{\gamma}_{\alpha\beta}{\cal
P}_{\gamma}- \frac{1}{24}c^{\alpha}c^{\beta}c^{\gamma}c^{\delta}
V^{\mu\nu}_{\alpha\beta}V^{\rho\sigma}_{\gamma\delta}F^{\lambda}_{\mu\rho}
{\cal P}_{\nu}{\cal P}_{\sigma}{\cal P}_{\lambda}.
\end{eqnarray}
Notice that BRST analysis for quadratic algebras of different special kinds 
including the case considered in \cite{SSvN}, was also given in \cite{DH}.
As to general nonlinear algebras of the form (\ref{j}), to our
knowledge, the problem of construction of a nilpotent the BRST
charge is open in this case.

In this paper
%, we investigate construction of the nilpotent BRST
%charge for nonlinear algebras (\ref{j})and
we find  some special restrictions on structure constants when the
nilpotent BRST charge (\ref{i2}) can be presented in the simplest
form including terms ${\cal Q}_1$ and ${\cal Q}_2$ only. The details
are given in \cite{BL}.

\section{BRST charge for generic nonlinear algebras}
Let us consider a theory with nonlinear algebras as described above (\ref{j}).
The structure constants $F^{\;\;\gamma}_{\alpha\beta}$ and
$V^{(k-1)\alpha_1...\alpha_k}_{\alpha\beta}\;(k=2,3,...,n)$ are
antisymmetric in lower indices and
$V^{(k-1)\alpha_1...\alpha_k}_{\alpha\beta}\;(k=2,3,...,n)$ are
totally symmetric in upper indices.

The Jacobi identities for these algebras have the form
\begin{eqnarray}\label{n2}
&&F^{\;\;\gamma}_{[\alpha\beta}F^{\;\;\delta}_{\lambda]\gamma}=0\;,\quad
F^{\;\;\rho}_{[\alpha\beta}V^{(1)\beta_1\beta_2}_{\lambda]\rho} +
V^{(1)\rho(\beta_1}_{[\alpha\beta} F^{\;\;\beta_2)}_{\lambda]\rho}=0
\;,\\
\nonumber
&&F^{\;\;\rho}_{[\alpha\beta}V^{(m)\beta_1...\beta_m\beta_{m+1}}_{\lambda]\rho}
+ V^{(m)\rho(\beta_1...\beta_m}_{[\alpha\beta}
F^{\;\;\beta_{m+1})}_{\lambda]\rho}+\\
\label{n3} &&+\sum_{k=1}^{m-1}C_{mk}
V^{(k)\rho(\beta_1...\beta_k}_{[\alpha\beta}
V^{(m-k)\beta_{k+1}...\beta_{m+1})}_{\lambda]\rho}
=0\; (m=2,3,...,n-1)\;, \\
\label{n4} &&\sum_{k=m-n+1}^{n-1}C_{mk}
V^{(k)\rho(\beta_1...\beta_{k}}_{[\alpha\beta}
V^{(m-k)\beta_{k+1}...\beta_{m+1})}_{\lambda]\rho}=0\;(m=n,...,2n-2)\;,
\end{eqnarray}
where
\begin{eqnarray}\label{Cmk}
 C_{mk}=\frac{(k+1)!(m-k+1)!}{(m+1)!}.
\end{eqnarray}
In Eqs. (\ref{n3}), (\ref{n4}) symmetrization includes  two sets of
symmetric indices. We assume that in the symmetrization only one
representative among equivalent ones obtained by permutation of
indices into these sets is presented.

Contribution of the first order in ghost fields $c^{\alpha}$,
 ${\cal Q}_1=c^{\alpha}T_{\alpha}$, defines the nilpotency equation
in the second order which has the solution
\begin{eqnarray}\label{ne2}
{\cal Q}_2=-\frac{1}{2}c^{\alpha}c^{\beta}
\Big(F^{\;\;\gamma}_{\alpha\beta}+ {\bar
V}^{\gamma\beta}_{\alpha\beta}T_{\beta}\Big){\cal P}_{\gamma}.
 \end{eqnarray}
Here the notation
\begin{eqnarray}\label{Vbar}
 {\bar V}^{\gamma\beta}_{\alpha\beta}=\sum_{k=1}^{n-1}
V^{(k)\alpha\beta\sigma_1...\sigma_{k-1}}_{\mu\nu}T_{\sigma_1}
\cdot\cdot\cdot T_{\sigma_{k-1}}
\end{eqnarray}
is used. Analyzing the nilpotency equation in the third order in ghost fields $c^{\alpha}$ we can find
that if the following  restrictions on
structure constants of the algebra (\ref{j})
\begin{eqnarray}\label{m8}
&&V^{(k)\sigma\beta_1\sigma_1...\sigma_{k-1}}_{[\alpha_1\alpha_2}
V^{(m-k)\sigma_{k}...\sigma_m}_{\alpha_3]\sigma}=0,\\
\nonumber &&k=1,2,...,n-1,\;\; m>k,\;\; m=2,3,...,2n-2.
\end{eqnarray}
are fulfilled, then the contribution to the BRST charge in the third order is equal to zero,
${\cal Q}_3=0$. Further analysis of the nilpotency equation in the forth order leads to the
following contribution to the BRST charge
\begin{eqnarray}
\label{n23} {\cal
Q}_4=-\frac{1}{24}c^{\alpha_1}c^{\alpha_2}c^{\alpha_3}c^{\alpha_4}
\Big(\bar{V}^{\sigma\beta_1}_{\alpha_1\alpha_2}
\bar{V}^{\beta_2\rho}_{\alpha_3\alpha_4}F^{\beta_3}_{\sigma\rho}+
2\bar{V}^{\sigma\beta_1}_{\alpha_1\alpha_2}
\tilde{V}^{\beta_2\rho}_{\alpha_3\alpha_4}F^{\beta_3}_{\sigma\rho}+
\tilde{V}^{\sigma\beta_1}_{\alpha_1\alpha_2}
\tilde{V}^{\beta_2\rho}_{\alpha_3\alpha_4}F^{\beta_3}_{\sigma\rho}\Big)
{\cal P}_{\beta_1}{\cal P}_{\beta_2}{\cal P}_{\beta_3},
\end{eqnarray}
where the notation
\begin{eqnarray}
\label{n17} \tilde{V}^{\alpha\beta}_{\mu\nu}=\sum_{k=1}^{n-1}
(k-1)V^{(k)\alpha\beta\sigma_1...\sigma_{k-1}}_{\mu\nu}T_{\sigma_1}
\cdot\cdot\cdot T_{\sigma_{k-1}}.
\end{eqnarray}
was used. In the case of quadratically nonlinear algebras ${\bar V}^{\gamma\beta}_{\alpha\beta}=V^{(1)\gamma\beta}_{\alpha\beta},
\tilde{V}^{\alpha\beta}_{\mu\nu}=0$ and from (\ref{n23}) it follows the result (\ref{i3})
in the forth order. The relations
\begin{eqnarray}
\label{n20} &&\bar{V}^{\sigma\beta_1}_{[\alpha_1\alpha_2}
\bar{V}^{\rho(\beta_2}_{\alpha_3\alpha_4]}F^{\beta_3)}_{\sigma\rho}=0,\quad
\bar{V}^{\sigma[\beta_1}_{[\alpha_1\alpha_2}
\tilde{V}^{\beta_2]\rho}_{\alpha_3\alpha_4]}F^{\beta_3}_{\sigma\rho}+
\bar{V}^{\sigma[\beta_1}_{[\alpha_1\alpha_2}
\tilde{V}^{\beta_3]\rho}_{\alpha_3\alpha_4]}F^{\beta_2}_{\sigma\rho}=0,\\
\label{n21} &&\bar{V}^{\sigma\beta_1}_{[\alpha_1\alpha_2}
\bar{V}^{\rho(\beta_2}_{\alpha_3\alpha_4]}
\bar{V}^{\beta_3\lambda)}_{\sigma\rho}=0,\quad
\bar{V}^{\sigma[\beta_1}_{[\alpha_1\alpha_2}
\tilde{V}^{\beta_2]\rho}_{\alpha_3\alpha_4]}
\bar{V}^{\beta_3\lambda}_{\sigma\rho}+cycle(\beta_2,\beta_3,\lambda)=0,
\\
&&\tilde{V}^{\sigma\beta_1}_{[\alpha_1\alpha_2}
\tilde{V}^{\rho(\beta_2}_{\alpha_3\alpha_4]}F^{\beta_3)}_{\sigma\rho}=0,\quad
\tilde{V}^{\sigma\beta_1}_{[\alpha_1\alpha_2}
\tilde{V}^{\rho(\beta_2}_{\alpha_3\alpha_4]}
\bar{V}^{\beta_3\lambda)}_{\sigma\rho}=0
\end{eqnarray}
derived from the Jacobi identities
(\ref{n2}) -- (\ref{n4}) and the restrictions (\ref{m8}) were used to
obtain the contribution (\ref{n23}).
We point out that the restrictions (\ref{m8}) lead to equalities
\begin{eqnarray}
\label{n22} \bar{V}^{\sigma\beta_1}_{[\alpha_1\alpha_2}
\bar{V}^{\beta_2\rho}_{\alpha_3\alpha_4]}
\bar{V}^{\beta_3\lambda}_{\sigma\rho}=0,\quad
\tilde{V}^{\sigma\beta_1}_{[\alpha_1\alpha_2}
\tilde{V}^{\beta_2\rho}_{\alpha_3\alpha_4]}
\bar{V}^{\beta_3\lambda}_{\sigma\rho}=0,\quad
\bar{V}^{\sigma[\beta_1}_{[\alpha_1\alpha_2}
\tilde{V}^{\beta_2]\rho}_{\alpha_3\alpha_4]}
\bar{V}^{\beta_3\lambda}_{\sigma\rho}=0.
\end{eqnarray}
If now we additionally assume the following restrictions on the
structure constants
\begin{eqnarray}
\label{m9}
&&V^{(k)\beta\beta_1\sigma_1...\sigma_{k-1}}_{[\alpha_1\alpha_2}
V^{(m-k)\beta_2\gamma\sigma_k...\sigma_{m-2}}_{\alpha_3\alpha_4]}
F^{\beta_3}_{\beta\gamma}=0,\\
\nonumber &&k=1,...,n-1,\quad m>k,\quad m=2,...,2n-2,
\end{eqnarray}
then we have
\begin{eqnarray}
\label{n25} \bar{V}^{\sigma\beta_1}_{[\alpha_1\alpha_2}
\bar{V}^{\beta_2\rho}_{\alpha_3\alpha_4]}F^{\beta_3}_{\sigma\rho}=0,\quad
\bar{V}^{\sigma[\beta_1}_{[\alpha_1\alpha_2}
\tilde{V}^{\beta_2]\rho}_{\alpha_3\alpha_4]}F^{\beta_3}_{\sigma\rho}=0,\quad
\tilde{V}^{\sigma\beta_1}_{[\alpha_1\alpha_2}
\tilde{V}^{\beta_2\rho}_{\alpha_3\alpha_4]}F^{\beta_3}_{\sigma\rho}=0
\end{eqnarray}
and as the result ${\cal Q}_4=0$.
Therefore  there exists a unique form of the nilpotent BRST
charge ${\cal Q}={\cal Q}_1+{\cal Q}_2$
if conditions (\ref{m8}), (\ref{m9}) are fulfilled.
 Although these conditions look like very restrictive,
there exist the interesting algebras where they are fulfilled. For
example, the conditions (\ref{m8}), (\ref{m9}) take place for
Zamolodchikov's $W_{3}$ algebra with central extension
 and for the higher spin algebras in AdS space
\cite{HSF}. Of course, there exist non-linear algebras for which the
conditions (\ref{m8}) and (\ref{m9}) are not fulfilled, e.g. these
relations are not valid for $so(N)$-extended superconformal algebras
with central extension \cite{CFT} (see also \cite{SSvN}).

\section{Summary}

In this paper we have studied a construction of the nilpotent
classical BRST charge for nonlinear algebras of the form (\ref{j})
which are characterized by the structure constants
$F^{\;\;\gamma}_{\alpha\beta},V^{(1)\alpha_1\alpha_2}_{\alpha\beta},...,
V^{(n-1)\alpha_1...\alpha_n}_{\alpha\beta}$.
We have proved that if the conditions (\ref{m8}) and (\ref{m9}) are satisfied
and a set of constraints $T_{\alpha}$ is linearly independent, the
BRST charge is given in the universal form ${\cal Q}={\cal Q}_1+{\cal Q}_2$.
Also we have proved that suitable quantities in terms of which one
can efficiently analyze general nonlinear algebras (\ref{j}) are
${\bar V}^{\mu\nu}_{\alpha\beta},{\tilde V}^{\mu\nu}_{\alpha\beta}$.

\section*{Acknowledgements}
The work was partially supported by the INTAS grant, project
INTAS-03-51-6346, the RFBR grant, project No.\ 06-02-16346, grant
for LRSS, project No.\ 4489.2006.2, the DFG grant, project No.\ 436
RUS 113/669/0-3 and joint RFBR-DFG grant, project No.\ 06-02-04012.

\end{document}